\documentclass[12pt]{article}                              
\usepackage{amsmath,amsthm,amsfonts,amscd,eucal,latexsym,amssymb}
\newlength{\dinwidth}
\newlength{\dinmargin}
\newsymbol\rest 1316         

\numberwithin{equation}{section}

\begin{document}
\newtheorem{theorem}{Theorem}[section]
\newtheorem{proposition}[theorem]{Proposition}
\newtheorem{lemma}[theorem]{Lemma}
\theoremstyle{definition}
\newtheorem{definition}[theorem]{Definition}
\theoremstyle{remark}
\newtheorem{ack}{Acknowledgement} \renewcommand{\theack}{}
%
\noindent
\begin{center}
{ \Large \bf
        On Generalizations of the Spectrum
        Condition\,\footnote{Contributed to the proceedings of
        ``Mathematical Physics in Mathematics and Physics'', held in
        Siena, June 20-25, 2000, in honour of Sergio Doplicher and
        John E.\ Roberts}}
\\[30pt]
{\large \sc  Rainer Verch}
\\[20pt]
                 Institut f\"ur Theoretische Physik,\\[4pt]
                 Universit\"at G\"ottingen,\\[4pt]
                 Bunsenstr.\ 9,\\[4pt]
                 D-37073 G\"ottingen, Germany\\[4pt]
                 e-mail:
 verch$@$theorie.physik.uni-goettingen.de\\[10pt]
{\it Dedicated to Sergio Doplicher and John E.\ Roberts\\
on the occasion of  their 60th birthdays}
\end{center}
${}$\\[18pt]
{\small {\bf Abstract. }
It is well known that the spectrum condition, i.e.\ the positivity of
the energy in every inertial coordinate system, is one of the central
conceptual ingedients in model-independent approaches to relativistic
quantum field theory. When one attempts to formulate quantum field
theory in a model-independent manner on a curved background spacetime,
it is not immediately clear which concepts replace the spectrum
condition.
The present work is devoted to reviewing facets of this situation,
thereby focussing on one particular approach that attempts
to generalize the notion of energy-momentum spectrum by the notion of
``wavefront set'', which may be seen as an asymptotic high-frequency
part of the spectrum.}

\section{Introduction}
The relativistic spectrum condition, stating that in relativistic
quantum field theory the spectrum of the energy should be positive in
all inertial Lorentz-frames, is one of the basic ingredients in all
model-independent approaches to quantum field theory, and together
with the principle of locality, it is responsible for remarkable
results of the model-independent approach, such as the Reeh-Schlieder
Theorem, the PCT-Theorem, and the connection between spin and
statistics \cite{Jos,SW,Haag}.
 Moreover, more recent developments have clearly indicated that there is in
quantum field theory a deep and subtle connection between the
Tomita-Takesaki modular objects and the spectrum condition. This is a
very active line of research, and promises to provide new and
interesting insights concerning the operator algebraic structure of
quantum field theory. The reader is referred to the recent review by
Borchers \cite{Bor.Rev} on these matters.
\par
Stimulated by developments during, roughly, the past three decades, it has
been realized that quantum field theory in curved spacetime (QFT in
CST) is a subject that promises to have physical relevance (see
\cite{Ful,WaldII} as general references). The major
impetus came from Hawking's theoretical arguments for particle
emission by black holes \cite{Haw} derived in the framework of QFT in
CST (cf.\ also \cite{FH}). 
There are other phenomena which also belong to the area of QFT in
CST, like the Casimir effect \cite{Cas,KayCas,Ful}, whose experimental
verification has 
recently reached an astonishing degree of accuracy
\cite{Caseff}. As regards relevance to cosmology, there are
suggestions that by QFT in CST methods one may account for the
recently observed accelerated expansion of the universe \cite{ParRav}.
Furthermore, one may view QFT in CST as a preliminary, semiclassical
approach to quantum effects in gravitation, hoping that the insights
gained from QFT in CST may provide some guideline at least towards
rudiments of that much
sought for theory of quantum gravity.
\par   
At any rate, there is reason enough to consider the mathematical and
conceptual foundations of QFT in CST a subject worthy of
interest. When embarking on that subject, one notices right at the
beginning that on a spacetime manifold, isometry groups are
generically absent, and so the usual, flat space version of the
spectrum condition can obviously not be formulated. In connection to
this circumstance, there is no natural candidate for a vacuum state
and, in turn, there is no natural choice of a set of physical states.  
\par
A guideline to finding a replacement of the spectrum condition for QFT
in CST originated from the study of free fields. An important initial
step was the approach by Wald \cite{Wal78} to defining the expectation
value of the energy-momentum tensor for states of a free scalar field
whose two-point functions are of Hadamard form. Following a number of
investigations (see, e.g.\ \cite{Ful,WaldII} 
and references therein),
it was realized that those Hadamard states are a good choice of a set
of physical states, comprising e.g.\ the set of so-called ``adiabatic
vacuum states'' that had been proposed by Parker for cosmological
model spacetimes (cf.\ \cite{Par,LRob,Jun}). A major advance in the
understanding of why Hadamard states are in a sense ``vacuum-like''
as regards their spectral behaviour was reached in Radzikowski's PhD
thesis \cite{Rad1}. Radzikowski showed that for a free scalar field 
the Hadamard form of a two-point function can be characterized,
in a one-to-one fashion, by a
specific form of the wavefront set of that two-point function. This
specific form can naturally be read as the generally covariant version
of the form of the wavefront set of a vacuum two-point function in
flat space; it is in a certain way asymmetric and this signifies a
high-frequency, short-distance remnant of the spectrum condition
(namely, the conicity of the energy-momentum spectrum).
\par
The characterization of Hadamard form in terms of conditions on the
form of the wavefront set has another advantadge: The Hadamard form
can only be prescribed for fields whose dynamics is governed by a free
wave-equation, while conditions on the wavefront set of $n$-point
functions generalizing the form of the wavefront set of $n$-point
vacuum expectation values in flat spacetime can be formulated also for
arbitrary, interacting field theories. This route has been taken in
\cite{BFK}, where suggestions for conditions on the wavefront set of
$n$-point functions of scalar fields on curved spacetime have been
made which are to be viewed as generalizing the flat space spectrum
condition.
With the help of these conditions, which are now referred to as
``microlocal spectrum condition'', abbreviated $\mu$SC, it is possible
to define Wick-powers, and their time-ordered products, of free
quantum field theories on curved spacetime, and moreover, to modify
the ``causal perturbation theory'' approach by Epstein and Glaser so
as to obtain a causal, local, perturbative construction of $P(\phi)_4$
scalar quantum field theories on general (globally hyperbolic) curved
spacetime. These very interesting results have been obtained recently by
Brunetti and Fredenhagen \cite{BrunFr}.
\par
However, the notion of wavefront set, and therefore, concepts of a
microlocal spectrum condition, so far required the formulation of a
quantum field theory in terms of quantum fields, or ``Wightman
distributions''. From a purely operator-algebraic point of view, it
appears highly desirable to have a generalized notion of the wavefront
set concept which is directly applicable to algebraic quantum field
theory. We have made a first attempt in that direction in
\cite{Ver.acs}, where we defined the ``asymptotic correlation
spectrum'' of a state, which may be viewed as the generalization of
the wavefront set to the operator-algebraic framework of quantum field
theory. Much of the present work (Sections 4 and 5) is devoted to this topic.
\par
This work is organized as follows. In Sec.\ 2 we recall afew basic
facts about the spectrum condition in flat spacetime. Sec.\ 3 is
concerned with a summary of aspects of quantum field theory in curved
spacetime. It begins with a collection of some general facts in Sec.\
3.1. In Sec.\ 3.2, the free scalar field is considered. The microlocal
spectrum condition will be discussed in Sec.\ 3.3. In Sec.\ 4 we
present the counterpart of the wavefront set concept in algebraic
quantum field theory, the ``asymptotic correlation spectrum'' of a
state. Most of the material in that section is taken from
\cite{Ver.acs}. In Sec.\ 5 we discuss some generalizations of the
asymptotic correlation spectrum to quantum field theory in curved spacetime. 
\section{Spectrum Condition}
Our discussion will be staged in the framework of the
operator-algebraic approach to quantum field theory, and so we begin
by recalling the basic structures of that framework. This will
be done, to start with, for Minkowski-space $\mathbb{R}^4$ as underlying
spacetime (of dimension = 4, but the setting is easily generalized to
any dimension $\ge 2$).
\par
In the operator-algebraic approach a quantum field theory is described
by a collection of objects $(\{\mathcal{A(O)}\}_{\mathcal{O} \subset
  \mathbb{R}^4}, (\alpha_x)_{x \in \mathbb{R}^4}, \omega^0)$, where it
is assumed that the following properties hold:
\begin{itemize}
\item[(a)] $\{\mathcal{A(O)}\}_{\mathcal{O} \subset
  \mathbb{R}^4}$ is a local net of $C^*$-algebras indexed by the
\textit{bounded open regions} $\mathcal{O}$ in $\mathbb{R}^4$, i.e.\
all $\mathcal{A(O)}$ are $C^*$-algebras containing a common unit
element, and the conditions of
\begin{itemize}
  \item isotony: $\mathcal{O}_1 \subset \mathcal{O}_2 \Rightarrow
    \mathcal{A(O}_1) \subset \mathcal{A(O}_2)$, and
  \item  locality: $ \mathcal{O}_1 \subset \mathcal{O}_2^{\perp}
    \Rightarrow A_1A_2 = A_2A_1\ \ \forall\ A_j \in \mathcal{A(O}_j) $
\end{itemize}
are fulfilled. Here $\mathcal{O}_2^{\perp}$ denotes the causal
complement set of $\mathcal{O}_2$ in the underlying spacetime (here,
Minkowski-space). 
\item[(b)] $(\alpha_x)_{x \in \mathbb{R}^4}$ is an automorphism group
  acting on the local net, i.e.\ the $\alpha_x$ are automorphisms of
  $\mathcal{A}$, the smallest $C^*$-algebra generated by all the
  $\mathcal{A(O)}$, and there holds $\alpha_x\alpha_y = \alpha_{x +
    y}$ and 
$$ \alpha_x(\mathcal{A(O)}) = \mathcal{A(O} + x)\,,$$
expressing that the $\alpha_x$ act covariantly as translations.
\item[(c)] $\omega^0$ is a vacuum state, i.e.\ $\omega^0$ is a state
  on $\mathcal{A}$ so that $x \mapsto \omega^0(A\alpha_x(B)C)$ is
  continuous for all $A,B,C \in \mathcal{A}$, and moreover, for all $f
  \in \mathcal{S}(\mathbb{R}^4)$ whose Fourier-transforms
  $\widehat{f}$ have support outside of the closed forward lightcone
  $\overline{V}_+$ \footnote{where $V_+ = \{x \in \mathbb{R}^4:
    (x^0)^2 - (x^1)^2 - (x^2)^2 - (x^3)^2 > 0,\ x^0 > 0\}.$}, it holds that
\begin{equation}
\label{speccond}
\int f(x)\,\omega^0(A^*\alpha_x(A))\,d^4x = 0 \quad \forall\ A \in
\mathcal{A}\,.
\end{equation}
\end{itemize}
One may then consider the GNS-representation
$(\mathcal{H}^0,\mathcal{\pi}^0,\Omega^0)$ of $\mathcal{A}$
corresponding to the vacuum state $\omega^0$. The von Neumann algebras
$\pi^0(\mathcal{A(O)})''$ will be denoted by $\mathcal{R(O)}$. They
contain all the observables of the underlying quantum field theory
which can be measured at times and locations in the spacetime region
$\mathcal{O}$. Since $\omega^0$ is a vacuum state, it follows that
there is a continuous unitary group $(U(x))_{x \in \mathbb{R}^4}$
implementing $(\alpha_x)_{x \in \mathbb{R}^4}$ in the
GNS-representation $\pi^0$, and the GNS-vector $\Omega^0$ is left
invariant under the action of $U(x)$, as can be deduced from
\eqref{speccond}. Moreover, \eqref{speccond} implies that the unitary
group $(U(x))_{x \in \mathbb{R}^4}$ fulfills the spectrum
condition. And this means that, if $(P_{\mu}) = (P_0,P_1,P_2,P_3)$
denote the generators of the unitary group, so that $U(x) =
\textrm{e}^{i P_{\mu}x^{\mu}}$, then $(P_0)^2 \ge (P_1)^2 + (P_2)^2 +
(P_3)^2$. In other words, the joint spectrum of the $P_{\mu}$ is
contained in $\overline{V}_+$. The vacuum state $\omega^0$ is
therefore a translation-invariant state of lowest energy; the
existence of such a state may be interpreted as a stability property
of the dynamics governing the quantum field theory described by 
$(\{\mathcal{A(O)}\}_{\mathcal{O} \subset
  \mathbb{R}^4}, (\alpha_x)_{x \in \mathbb{R}^4}, \omega^0)$.
\par
The above stated conditions (a,b,c) may be viewed as minimal
conditions for the mathematical description of a quantum field theory
in the operator-algebraic framework. They are usually supplemented by
further conditions expressing additional properties of the quantum
field theory to be described. (See \cite{Haag} for ample discussion.)
One such condition is, e.g., Poincar\'e-covariance. Another condition
typically imposed is that $\omega^0$ be a pure state on $\mathcal{A}$,
which can be shown to be equivalent to asymptotic spacelike
clustering. A further condition is to strengthen weak continuity (with
respect to the vacuum folium) of $(\alpha_x)_{x \in \mathbb{R}^4}$ to
strong continuity, meaning that $||\,\alpha_x(A) - A\,|| \to 0$ for $x
\to 0$ holds for all $A \in \mathcal{A}$. (As has been pointed out in
\cite{BV1}, given the vacuum representation $\pi^0$ or any other
representation of $\mathcal{A}$ in which $(\alpha_x)_{x \in
  \mathbb{R}^4}$ acts weakly continuously, each $\mathcal{R(O)}$
contains a weakly dense subalgebra on which the action of
$(\alpha_x)_{x \in \mathbb{R}^4}$ is strongly continuous. Thus the
assumption of strong continuity for the translations doesn't appear to
be too restrictive.) 
\par
Under the above stated assumptions (a) and (b) together with strong
continuity of $(\alpha_x)_{x \in \mathbb{R}^4}$, Doplicher \cite{Dop}
proved that $\mathcal{A}$ admits a vacuum state $\omega^0$ if and only
if the spectral ideal $\mathcal{J} \subset \mathcal{A}$ is
proper. Here, the spectral ideal $\mathcal{J}$ is the left 
ideal in $\mathcal{A}$ generated by all $A$
having the property that $\int f(x)\,\alpha_x(A)\,d^4x = 0$ holds for
all $f \in \textrm{L}^1(\mathbb{R}^4)$ whose Fourier-transforms are
supported outside of $\overline{V}_+$. 
This characterization shows that the existence of vacuum states may be
seen as a property of the algebraic structure of the algebra of
observables $\mathcal{A}$ relative to the action of the translations.
\section{Quantum Field Theory on Curved Spacetime}
\subsection{Generalities}
The flat space spectrum condition clearly hinges upon the presence of
the translation group. When considering quantum field theories on a
curved spacetime, there is in general no counterpart of the
translation group, and in general, there is not even any time-symmetry
group.
These circumstances make it difficult to formulate what should be a
vacuum state for a quantum field theory on a curved spacetime. Even
worse, it is not even clear what the set of physical states should
be for quantum fields in a curved spacetime. Let us briefly recall how
the set $\mathcal{S}_{\textrm{phys}}$ of physical states may 
be determined from the vacuum state for a
given quantum field theory $(\{\mathcal{A(O)}\}_{\mathcal{O} \subset
  \mathbb{R}^4}, (\alpha_x)_{x \in \mathbb{R}^4}, \omega^0)$ on
Minkowski spacetime: Here, one usually takes
$\mathcal{S}_{\textrm{phys}}$ to consist of all states $\omega$ on
$\mathcal{A}$ which are locally normal to the vacuum state
$\omega^0$. This means that there is for each \textit{bounded} open
region $\mathcal{O} \subset \mathbb{R}^4$ a density matrix
$\rho_{\omega,\mathcal{O}}$ on 
$\mathcal{H}^0$ so that 
$$ \omega(A) = \textrm{Tr}(\rho_{\omega,\mathcal{O}}\cdot\pi^0(A)) $$
holds for all $A \in \mathcal{A(O)}$. That definition of ``physical
state'' is on one hand broad enough and allows ``charged'' states
(from which charge-carrying fields, originally not contained in
$\mathcal{A}$, can be constructed \cite{DR}), on the other hand it
avoids pathologies like states having infinite particle density which
would be highly unphysical \cite{Haag,Rob.Lec}.
\par
However, one can outline a basic a approach to the description of QFT
in CST in the operator-algebraic setting. (We should like to mention
that the formulation we are going to give here is patterned after several
precursors, as e.g.\ given in \cite{Dim,GLRV}. No claim of originality
is made at this stage.) To this end, a curved
spacetime will be modelled mathematically by a pair $(M,g)$ where $M$
is a 4-dimensional smooth manifold and $g$ a smooth metric on this
manifold of Lorentzian signature. To avoid any causal pathologies, we
shall assume that $(M,g)$ is globally hyperbolic. This means that the
manifold $M$ can be smoothly foliated in Cauchy-surfaces, where a
Cauchy-surface is a 3-dimensional sub-manifold which is intersected
exactly once by each inextendible, $g$-causal curve in $M$. We refer
to \cite{HE,WaldI} for further discussion and presentation of
examples; it should nevertheles be mentioned that the class of
globally hyperbolic spacetimes contains most of the spacetime models
thought to describe physically relevant situations (like
Robertson-Walker, de\,Sitter, Schwarzschild-Kruskal and, of course,
Minkowski-spacetime). Also, it is worth mentioning that global
hyperbolicity isn't related to the existence of spacetime isometries. 
\par
Assuming now that a globally hyperbolic spacetime
$(M,g)$ has been given to us, we formulate the basic mathematical
structure of a quantum field theory on this ``background spacetime'' 
as being described by a collection of objects 
$(\{\mathcal{A(O)}\}_{\mathcal{O} \subset M},(\alpha_{\gamma})_{\gamma
  \in G},$ $\mathcal{S}_{\textrm{phys}}^{0})$ with the following
properties:
\begin{itemize}
\item[(a')] $\{\mathcal{A(O)}\}_{\mathcal{O} \subset M}$ is a family
  assigning to each bounded (i.e., relatively compact) open region
  $\mathcal{O}$ in $M$ a $C^*$-algebra $\mathcal{A(O)}$ in such a way
  that all these algebras have a common unit element and so that the
  conditions of isotony and locality (which can be taken over
  literally from (a) above) are fulfilled.
\item[(b')]  $G$ denotes the group of proper, orthochronous
  isometries of the spacetime $(M,g)$, and $(\alpha_{\gamma})_{\gamma
    \in G}$ is a group of automorphisms of $\mathcal{A}$, the
  $C^*$-algebra generated by all $\mathcal{A(O)}$, with
  $\alpha_{\gamma_1}\alpha_{\gamma_2} = \alpha_{\gamma_1\gamma_2}$ and
  the covariance property
  $$ \alpha_{\gamma}(\mathcal{A(O)}) = \mathcal{A(}\gamma(\mathcal{O})) \,.$$
(There may frequently occur the case that $G$ contains just the
identical map; then the present condition is effectively void.)
\item[(c')] $\mathcal{S}_{\textrm{phys}}^0$ is a subset of the set of
  physical states, selected by a suitable generalization of the
  spectrum condition. The GNS-representations $\pi_1$ and $\pi_2$ of
  $\mathcal{A}$ corresponding to any pair of states $\omega_1,\omega_2
  \in \mathcal{S}_{\textrm{phys}^0}$ are  assumed to be locally
  quasi-equivalent (quasi-equivalent when restricted to $\mathcal{A(O)}$
  for any bounded region $\mathcal{O}$), and we suppose that $\gamma
  \mapsto \omega(A\alpha_{\gamma}(B)C)$ is continuous (for $\gamma$
  ranging over continuous parts of $G$).
\end{itemize}
It is clear that (a') and (b') are natural generalizations of (a) and
(b) above with similar meaning. Concerning (c'), what has essentially
been changed in comparison to (c) is that the existence of one
particular distinguished state has been replaced by a whole set of
states which are supposed to be dintinguished by a certain,
generalized form of the spectrum condition. The present
formulation of a mathematical framework is again to be viewed as, in a
sense, consisting of ``minimal'' requirements; as we shall see, to
make precise mathematical sense of ``generalized form of the spectrum
condition'' in the present abstract operator-algebraic setting 
one needs additional structure, in particular a generalization of (b')
is needed for the case that $G$ is very small. 
In the case that the operator-algebras $\pi(\mathcal{A(O)})''$ for
$\pi$ a GNS-representation of an $\omega \in
\mathcal{S}_{\textrm{phys}^0}$ are generated by quantum fields, we 
view the microlocal spectrum condition of \cite{BFK} as a candidate
for that generalized form of the spectrum condition. But we shall
follow the historical course of events and will first look at the
example of the free scalar field in the next subsection. Before turning
there, a word on the condition of local quasi-equivalence is in order.
The members $\omega$ in the set $\mathcal{S}_{\textrm{phys}}$ should
be locally normal to states $\omega^0$ in
$\mathcal{S}_{\textrm{phys}}^0$. However, as the members within
$\mathcal{S}_{\textrm{phys}}^0$ aren't further distinguished, 
consistency requires that each state $\omega$ on $\mathcal{A}$
which is locally normal to some $\omega^0 \in
\mathcal{S}_{\textrm{phys}}^0$ is also locally normal to any other 
$\hat{\omega}^0 \in\mathcal{S}_{\textrm{phys}}^0$. And this is just
equivalent to the condition of local quasi-equivalence formulated in
(c') above.
\subsection{Free scalar field and Hadamard states}
The following treatment of the free scalar field on a globally
hyperbolic spacetime $(M,g)$ is due to Dimock \cite{Dim}.
\par
The classical free scalar field obeys the field equation
\begin{equation}
\label{KGeqn}
 (\nabla^a \nabla_a + m^2) \varphi = 0\,,
\end{equation}
where $\varphi$ is a real-valued smooth function on $M$, $\nabla_a$
denotes the covariant derivative of the spacetime metric $g$, and $m
>0$ is a constant. On a globally hyperbolic spacetime, the
Cauchy-problem for the wave-equation \eqref{KGeqn} is well-posed, and
there is a unique pair of advanced/retarded solutions of
\eqref{KGeqn}, i.e.\ continuous linear operators $E^{\pm}:
C^{\infty}_0(M,\mathbb{R}) \to C^{\infty}(M,\mathbb{R})$ so that 
$E^{\pm}(\nabla^a\nabla_a + m^2)f = f = (\nabla^a\nabla_a +
m^2)E^{\pm}f$ holds for all $f \in C_0^{\infty}(M,\mathbb{R})$, and so
that the support of $E^{\pm}f$ is contained in the causal future/past
of $\textrm{supp}\,f$. The difference of the fundamental solutions
takes test-functions to solutions of \eqref{KGeqn} and is called the
propagator, or commonly also the commutator function. One can show
that on the quotient space $K =
C^{\infty}_0(M,\textrm{R})/\textrm{ker}(E)$ there is a symplectic form
$\sigma(\,.\,,\,.\,)$ given by
$$ \sigma([f],[h]) = \int_M f\cdot Eh\,d\mu $$
for all $[f],[h] \in K$, where we have denoted the quotient map
$C^{\infty}_0(M,\mathbb{R}) \to K$ by $f \mapsto [f]$ and the
metric-induced measure on $M$ by $d\mu$.
\par
Then it is standard to associate to the symplectic space $(K,\sigma)$
a $C^*$-algebra $\mathcal{A}[K,\sigma]$ generated by a family
$\{W([f]): [f] \in K\}$ of unitary elements satisfying the relations
$$ W([f])^* = W(-[f])\,, \quad W([f])W([h]) =
\textrm{e}^{-i\sigma([f],[h])/2}W([f] + [h])$$
for all $[f],[h] \in K$; these are called the canonical commutation
relations in Weyl-form.
A net $\{\mathcal{A(O)}\}_{\mathcal{O} \subset M}$ of local
$C^*$-algebras fulfilling the conditions of isotony and locality (a')
can then be obtained by setting
$$ \mathcal{A(O)} = C^*\textrm{-algebra\ generated\ by\ all\ }
W([f]),\ \textrm{supp}\,f \subset \mathcal{O}\,.$$
Moreover, one can show that this net of $C^*$-algebras fulfills also
condition (b'). See \cite{Dim,WaldII} for further discussion.
\par
In a next step, one has to make a choice of
$\mathcal{S}_{\textrm{phys}}^0$. First, one collects states that are
suitably regular and have a simple structure. For the Weyl-algebra
$\mathcal{A}[K,\sigma]$, a natural choice is to consider as candidates
the so-called quasi-free states. They take the form 
\begin{equation}
\label{qf}
\omega( W([f])) = \textrm{e}^{-\omega_2(f,f)/2} \quad \forall\ [f] \in K\,,
\end{equation}
where $\omega_2(\,.\,,\,.\,)$ is the two-point function of $\omega$,
defined by
 $$\omega_2(f,h) = \left.-\partial_{s}\partial_t\right|_{t
  = s = 0}\omega(W(s[f])W(t[h]))$$
 for all $f,h \in
C^{\infty}_0(M,\mathbb{R})$. In other words, quasi-free states $\omega$
are determined by their two-point functions $\omega_2$ via \eqref{qf}.
To restrict attention to quasi-free states when specifying conditions
for the initial set $\mathcal{S}_{\textrm{phys}}^0$ of physical states
thus constitutes a considerable simplification as now one needs only 
impose conditions on the two-point functions. The question is, then,
what the two-point function $\omega_2$ of a physical state of the free scalar
on a curved spacetime field should look like. As a technical condition
it seems natural to assume that $\omega_2$ is a distribution. Above
that, further input is required. The suggestion by Wald \cite{Wal78}
was that two-point functions of physical states should have Hadamard
form. We won't pause to discuss the motivations for that since this
has been done in some depth in the literature
\cite{Ful,WaldII}. However, we give a sketch of the definition of
Hadamard form. One says that $\omega_2$ is of Hadamard form if
$$  
  \omega_2(f,h) = \lim_{\varepsilon \to 0}\,\int_{M \times M}
  (G_{\varepsilon}(x,y) + H_{\omega}(x,y))f(x)h(y)
  \,d\mu(x)\,d\mu(y)\,,$$
where $H_{\omega} \in C^{\infty}(M \times M)$ contains the
dependence on the state  while the singular part, represented by
$\lim_{\varepsilon \to 0}\,G_{\varepsilon}$ is the same for all
Hadamard states and given -- qualitatively -- by
\begin{equation}
\label{Hadform}
G_{\varepsilon}(x,y) = \frac{U(x,y)}{s(x,y) + i\varepsilon(x,y)} +
V(x,y) \textrm{ln}(s(x,y) + i\varepsilon(x,y))\,.
\end{equation}  
Here, $s(x,y)$ denotes the square of the geodesic distance from $x$ to
$y$, $\varepsilon(x,y)$ is of order $\varepsilon$ and has
positive/negative sign according if $x$ lies in the future/past of
$y$, and $U$ and $V$ are smooth functions which are determined by the
wave-operator $(\nabla^a\nabla_a + m^2)$ by means of the so-called
Hadamard recursion relations (see \cite{Fri,Gun} for their modern
formulation as well as references to the original works by
Hadamard). This definition is only qualitative since $s(x,y)$ (and
likewise, $U(x,y)$ and $V(x,y)$) need not be defined globally for all
$x,y \in M$, and in fact it took some time until a completely
satisfactory definition of Hadamard form was first reached at in \cite{KW}.
\par
Since for all Hadamard forms their singular parts are identical, the
difference of any pair of Hadamard forms $\omega_2$ and
$\hat{\omega}_2$ is given by the smooth integral kernel $H_{\omega}
- H_{\hat{\omega}}$. One can show that this fact is sufficient in
order that the GNS-representations $\pi$ and $\hat{\pi}$ corresponding
to any pair of quasifree states $\omega$ and $\hat{\omega}$ on
$\mathcal{A}[K,\sigma]$ having two-point functions of Hadamard form
-- such states 
will henceforth be called Hadamard states --  are locally
quasi-equivalent \cite{Ver1}. Thus, if one chooses for the free scalar
field on a globally hyperbolic spacetime as initial
collection of physical states $\mathcal{S}_{\textrm{phys}}^0$ the set of
Hadamard states, then condition (c') above is clearly fulfilled.
\par
While the Hadamard condition thus appears as a reasonable selection criterion 
for physical states of the free scalar field (which may similarly be
generalized to other fields obeying linear wave equations, cf.\ e.g.\
\cite{SV2} and references cited therein), it is not
immediately clear what the Hadamard form has to do with with a
``suitable generalization of the spectrum condition'', which we had
desired above to distinguish the set
$\mathcal{S}_{\textrm{phys}}^0$. In particular, since Hadamard forms
are only definable with respect to linear wave-equations, it is at
this stage not at all evident how to generalize the Hadamard form
criterion to more general quantum field theories. These points have
been significantly clarified in the PhD thesis of Radzikowski
\cite{Rad1} who noticed that the wavefront set of a Hadamard form
assumes a distinguished shape.  
\subsection{Wavefront sets and microlocal spectrum condition}
In order to present and discuss Radzikowski's findings, we fist have
to introduce the notion of the wavefront set of a scalar distribution.
There are several equivalent definitions that one can give, but
perhaps the simplest approach is the one we take here. See
\cite{Hor1} for further discussion.
\par
Let $n \in \mathbb{N}$ and $v \in \mathcal{D}'(\mathbb{R}^n)$. One calls
$(x,k) \in \mathbb{R}^n 
\times (\mathbb{R}^n\backslash\{0\})$ a {\it regular directed point} for $v$
if there are $\chi \in \mathcal{D}(\mathbb{R}^n)$ 
with $\chi(x) \ne 0$, and a conical
open neighbourhood $\Gamma$ of $k$ in $\mathbb{R}^n \backslash\{0\}$
[i.e.\ $\Gamma$ 
is an open neighbourhood of $k$, and $k \in \Gamma \Leftrightarrow \mu k \in
\Gamma$ $\forall\, \mu > 0$], such that
$$ \sup_{\tilde{k} \in \Gamma}\,(1 + |\tilde{k}|)^N |\widehat{\chi
  v}(\tilde{k})| \le C_N < \infty $$
holds for all $N \in \mathbb{N}$, where $\widehat{\chi v}$ denotes the
Fourier transform of the distribution $\chi\cdot v$.
\begin{definition}
$\textrm{WF}(v)$, the {\em wavefront set} of $v \in
\mathcal{D}'(\mathbb{R}^n)$,  is
defined as the complement in $\mathbb{R}^n \times (\mathbb{R}^n
\backslash\{0\})$  of
the set of all regular directed points for $v$.
\end{definition}
Thus, $\text{WF}(v)$ consists of pairs $(x,k)$ of points $x$ in
configuration space, and $k$ in Fourier space, so that the Fourier
transform of $\chi\cdot v$ isn't rapidly decaying along the direction
$k$ for large $|k|$, no matter how closely $\chi$ is concentrated
around $x$.

If $\phi : U \to U'$ is a diffeomorphism between open subsets of
$\mathbb{R}^n$, and $v \in \mathcal{D}'(U)$,
 then it holds that $\textrm{WF}(\phi_*v) =
({}^tD\phi)^{-1} \textrm{WF}(v)$ where ${}^tD\phi$ denotes the transpose of the
tangent map (or differential) of $\phi$, with $({}^tD\phi)^{-1}(x,k) =
(\phi(x),({}^tD\phi)^{-1}\cdot k)$ for all $(x,k) \in \textrm{WF}(v)$ and
$\phi_*v(f) = v(f \circ\phi)$, $f \in \mathcal{D}(U')$. This
transformation behaviour of the wavefront set allows it to define the
wavefront set $\textrm{WF}(v)$ of a scalar distribution
$v \in \mathcal{D}'(X)$ on any
$n$-dimensional manifold $X$ [as usual, we take manifolds to be
Hausdorff, connected, 2nd countable, $C^{\infty}$ and without boundary]
by using coordinates: Let $\kappa : U \to \mathbb{R}^n$ be a coordinate system
around a point $q \in X$. Then the inverse dual tangent map is an isomorphism
$({}^tD\kappa)^{-1} : T_q^*X \to \mathbb{R}^n$. We will use the notational
convention $(q,\xi) \in T^*X \Leftrightarrow \xi \in 
  T_q^*X$. Then let $(q,\xi) \in {\rm T}^*X\backslash\{0\}$ and
$(x,k) := ({}^tD\kappa)^{-1}(q,\xi) =
(\kappa(q),({}^tD\kappa)^{-1}\cdot \xi)$,  so that $(x,k)$
is in $\mathbb{R}^n \times(\mathbb{R}^n \backslash\{0\})$.
\begin{definition}
We define $\textrm{WF}(v)$ by saying that $(q,\xi) \in \textrm{WF}(v)$ iff
$(x,k) \in \textrm{WF}(\kappa^*v)$ where $\kappa^*v$ is the chart expression of
$v$.
\end{definition}
Owing to the transformation properties of the wavefront set under
local diffeomorphisms one can see that this definition is independent
of the choice of the chart $\kappa$, and moreover, $\text{WF}(v)$ is a
subset of $T^*X \backslash\{0\}$, the cotangent bundle without the
zero section.

It is straightforward to deduce from the definition that
$$
 \textrm{WF}(Av) \subset \textrm{WF}(v)\,, \quad v \in \mathcal{D}'(X)\,,
$$
for any partial differential operator $A$ with smooth coefficients. 
(This generalizes to
pseudodifferential operators $A$.) It is also worth noting that
$\textrm{WF}(v)$ is a closed conic subset of $T^*X \backslash\{0\}$ 
where conic means $(q,\xi) \in \textrm{WF}(v) \Leftrightarrow (q,\mu \xi)
\in \textrm{WF}(v)$ $\forall\,\mu > 0$. Another important property is the
following: Denote by $p_{M^*}$ the base projection of $T^*X$,
i.e.\ $p_{M^*}: (q,\xi) \mapsto q$. Then for all $v \in \mathcal{D}'(X)$ there
holds
\begin{equation}
 p_{X^*}\textrm{WF}(v) = \textrm{sing\,supp}\,v
\end{equation}
where sing\,supp\,$v$ is the {\em singular support} of $v$.
\begin{definition}
\label{D-3}
For $v \in \mathcal{D}'(X)$, $\textrm{sing\,supp}\,v$ is defined as the
complement of all points $q \in X$ for which there is an open
neighbourhood $U$ and a smooth $n$-form $\alpha_U$ on $U$ so that
$$ v(h) = \int_U h\cdot \alpha_U\quad \textrm{for\ all}\ \ h \in \mathcal{D}(U)\,.$$
\end{definition}
In other words, $v$ is given by an integral over a smooth $n$-form
exactly if $\textrm{WF}(v)$ is empty.
\par
Now let $(M,g)$ be a globally hyperbolic spacetime. Then define the
set of ``null-covectors'' 
\begin{equation}
\label{nullcov}
   \mathcal{N} =\{(q,\xi) \in T^*M : g^{\sigma \rho}(q)
   \xi_{\sigma}\xi_{\rho} = 0\}\,.
\end{equation}
The spacetime possesses a time-orientation, i.e.\ there is on $M$ a
vector field $w$ which is timelike, hence everywhere non-zero, and, by
definition, future pointing. With its help one can introduce the
following two disjoint future/past-oriented parts of $\mathcal{N}$,
$$
    \mathcal{N}_{\pm} = \{(q,\xi) \in \mathcal{N}: \pm \xi(w) >
    0\}\,.$$
On the set $\mathcal{N}$ one can introduce an equivalence relation as
follows:
\begin{definition}
One defines
  $$ (q,\xi) \sim (q',\xi') $$
if there is an affinely parametrized lightlike geodesic $\gamma$ with
$\gamma(t) = q$, $\gamma(t') = q'$ and
$$ g^{\sigma \rho}(q)\xi_{\rho} =
(\left. \mbox{$\frac{d}{ds}$}\right|_{s = t}\gamma(s))^{\sigma}\,,
\quad g^{\sigma \rho}(q')\xi'_{\rho} =
 (\left. \mbox{$\frac{d}{ds}$}\right|_{s = t'}\gamma(s))^{\sigma}\,.$$
\end{definition}
In other words, $\xi$ and $\xi'$ are co-parallel to the lightlike
geodesic $\gamma$ connecting the base points $q$ and $q'$, and
therefore $\xi$ and $\xi'$ are parallel transports of each other along
that geodesic.
\par
Equipped with that notation, we can now formulate Radzikowski's
result, which rests to some extent on previous work by Duistermaat
and H\"ormander \cite{DH}.
\begin{theorem}[Radzikowski]
Let $\omega_2 \in \mathcal{D}'(M \times M)$ be the two-point function
of a state on the Weyl-algebra $\mathcal{A}[K,\sigma]$ of the free
scalar field on the globally hyperbolic spacetime $(M,g)$. Then
$\omega_2$ is of Hadamard form if and only if
\begin{equation}
\label{charhadform}
 {\rm WF}(\omega_2) = \{(q,\xi;q',\xi') \in \mathcal{N}_- \times
 \mathcal{N}_+ : (q,\xi) \sim (q',-\xi')\}\,.
\end{equation}
\end{theorem}
What is so attractive about this characterization of Hadamard forms?
First, \eqref{charhadform} is just the generally covariant
generalization of the form of the wavefront set for the two-point
function of the Klein-Gordon field's vacuum state in flat Minkowski
spacetime. Secondly, it expresses an asymptotic high-frequency remnant
of the spectrum condition, which in flat spacetime may be expressed as
a restriction on the Fourier-space support of the translation group as
in condition (c) of Section 2. We will make this somewhat more
precise in the next section. Moreover, and quite importantly, a
condition of the type \eqref{charhadform} can be generalized to other
than just free quantum fields. A significant step in this direction
has been taken by Brunetti, Fredenhagen and K\"ohler \cite{BFK}. We
briefly sketch their ``microlocal spectrum conditon'' ($\mu$SC).
\par
Assume that $(M,g)$ is, as before, a globally hyperbolic spacetime,
and let $\mathcal{B}_M$ denote the Borchers-algebra over the manifold
$M$. That is, $\mathcal{B}_M$ is the free tensor-algebra of scalar
test-functions, in symbols
$\mathcal{B}_M = \mathbb{C} \oplus_{n \in \mathbb{N}}(\otimes^n
\mathcal{D}(M))$; an algebraic structure can be defined on
$\mathcal{B}_M$ is a canonical way \cite{Bor1}. A state $\omega$ on
$\mathcal{B}_M$ is a positive linear functional which is uniquely
specified by a sequence $(\omega_m)_{m \ge 0}$ where $\omega_0 \in
\mathbb{C}$ and the $m$-point functions (or $m$-point distributions)
$\omega_m \in (\otimes^m\mathcal{D}(M))'$ are the restrictions of
$\omega$ to $\otimes^m \mathcal{D}(M)$. The approach by Brunetti,
Fredenhagen and K\"ohler is to impose restrictions on
$\textrm{WF}(\omega_m)$ which are to viewed as generalizations of the
flat space spectrum condition. For the vacuum state $\omega$ in flat
spacetime, the spectrum condition amounts to restricting the support
of the Fourier-transform of $\omega_m$ for each $m$ in a specific way,
encoding that the energy-momentum spectrum is ``conic'' and that
$\omega$ is translation-invariant \cite{SW,Bor1}. The
terminology ``microlocal spectrum condition'' refers to the fact that
the wavefront set is the microlocal version of the Fourier-space
support of a distribution, in the sense that the distribution is 
localized around a point and support properties of its
Fourier-transform are replaced by rapid decay properties. This notion
is, as the above stated transformation properties of the wavefront set
show, independent of the chosen coordinate system, while the support
of a distribution's Fourier-transform is a manifestly
coordinate-dependent concept. This indicates once more the utility of
the notion of wavefront set for generalizing the spectrum condition to
quantum field theory in curved spacetime.
\par
To eventually formulate the microlocal spectrum condition, it is
necessary to introduce further notation. Let $G$ be a non-directed 
graph with $n$ vertices $\{v_1,\ldots,v_n\}$ and a collection of
connecting edges $E_G = \{e_1,\ldots,e_N\}$. More precisely, a
directed edge $\vec{e}_{ij} = \langle v_i,e,v_j\rangle$ is an edge
connecting the source-vertex $v_i$ to the range-vertex $v_j$, and to
say that the graph $G$ is non-directed means that, if $\vec{e}_{ij}$
is contained in $E_G$, then also its opposite directed edge,
$(\vec{e}_{ij})' \equiv \vec{e}_{ji}$ is contained in $E_G$. Note that
there may be several different edges in $E_G$ connecting the same
source- and range-index pair, and it is also allowed that there are
isolated vertices in $\{v_1,\ldots,v_n\}$ which aren't source- or
range-vertices of any directed edge in $E_G$. 
Now an {\it immersion} of a non-directed graph $G$ (with vertices
$\{v_1,\ldots,v_n\}$) into the spacetime $(M,g)$ is defined as a map
$\iota (\,.\,)$ with the following properties: (1) to each vertex
$v_i$, it assigns a point $p_i = \iota (v_i)$ in $M$, (2) to each
directed edge $\vec{e}_{ij} \in E_G$ it assigns a covector
$(p_i,\xi) = \iota(\vec{e}_{ij})\in T^*_{x_i}M$, with $p_i = \iota
(v_i)$, together with 
a smooth curve $\gamma_{ij}$ connecting $p_i$ and $p_j = \iota
(v_j)$, (3) for $(\vec{e}_{ij})' = \vec{e}_{ji}$, and $(p_j,\xi') =
\iota (\vec{e}_{ji})$, it is required that $\gamma_{ji} =
\gamma_{ij}$ and that $\xi'$ is the parallel transport of $-\xi$
along $\gamma_{ij}$, (4) if $i > j$, then (the dual of) the covector
$(p_i,\xi) = \iota (\vec{e}_{ij})$ is causal and future-directed.
\par
With this notation, the microlocal spectrum condition of \cite{BFK}
reads as follows.
\begin{definition}
A state $\omega$ on $\mathcal{B}_M$ with $m$-point functions
$\omega_m$ is said to satisfy the microlocal spectrum condition
($\mu$SC) iff
$$ \textrm{WF}(\omega_m) \subset \Gamma_m \quad \textrm{for\ all}\
m\,,$$
where $\Gamma_m$ is defined as the set of all
$(p_1,\xi_1;\ldots;p_m,\xi_m) \in (T^*M^m)\backslash \{0\}$ so that
there exists a non-directed graph $G$ with $m$ vertices
$\{v_1,\ldots,v_m\}$ together with an immersion $\iota (\,.\,)$ into
$(M,g)$ having the properties
$$ p_i = \iota (v_i) \quad \textrm{and} \quad (p_i,\xi_i) = \sum_j
\iota (\vec{e}_{ij})\,.$$
\end{definition}
We remark that $\Gamma_m$ is a covariant generalization of the set
bounding the Fourier-space support of the Wightman $m$-point
functions in Minkowski-spacetime; the condition that (the dual of)
$\xi_m$ be future-pointing and causal corresponds to the spectrum
condition while the requirement that $(p_i,\xi_i) = \sum_j
\iota(\vec{e}_{ij})$ is the microlocal remnant of translation
invariance. We refer to \cite{BFK} for more discussion on this point. 
\par
The set $\Gamma_m$ may be quite ``large'' as regards the relative
position of the base-points $p_1,\ldots,p_m$, since the connecting
curves $\gamma_{ij}$ appearing in the definition of a graph-immersion
are only required to be smooth. In this respect, it is at present not
completely clear if the definition of a graph-immersion shouldn't be
more restrictive. In \cite{BrunFr}, a graph-immersion is defined in a more
restrictive manner, where the $\gamma_{ij}$ are required to be
lightlike geodesics with dual tangent $\xi_i$ at $p_i$. With this
modified, more restrictive definition of graph-immersions and
correspondingly, of $\Gamma_m$, it is shown in \cite{BrunFr} that the
$m$-point functions of quasifree Hadamard states for the free scalar
field fulfill the bound $\textrm{WF}(\omega_m) \subset
\Gamma_m$. This property is an important technical tool for the local
perturbative construction of $P(\phi)_4$ theories in globally
hyperbolic curved spacetimes developed in \cite{BrunFr}.
\section{Asymptotic Correlation Spectrum}
The previous considerations have shown that the wavefront set of the
$m$-point correlation functions $\omega_m$ for states on the Borchers
algebra is a very useful concept in order to formulate generalized
versions of the spectrum condition for quantum field theory in curved
spacetime. However, the description of a quantum field theory in terms
of $m$-point correlation functions, or equivalently, in terms of
quantum fields (operator-valued distributions) is not an intrinsic
concept from the point of view of algebraic quantum field theory as
outlined in Section 2. One would like to generalize the concept of
wavefront set in such a way that it becomes an intrinsic notion within
the framework of algebraic quantum field theory, say, in Minkowski
spacetime to start with, using only the structural assumptions (a) and
(b) of Sec.\ 2 which are also prerequisite to the spectrum
condition. Such an algebraic variant of the wavefront set would then
be an invariant of a state, or of a set of states, in a similar manner
as the spectrum condition, and would be independent of the the various
choices of different quantum fields that one may have to generate the
same net of von Neumann algebras $\{\mathcal{R(O)}\}_{\mathcal{O}
  \subset \mathbb{R}^4}$.
\par
We begin our discussion by collecting the relevant assumptions.
Suppose that $\{\mathcal{A(O)}\}_{\mathcal{O}\subset \mathbb{R}^n}$ is
an isotonous family of $*$-algebras indexed by the bounded
open regions in $\mathbb{R}^n$. That is to say, to each bounded open
region $\mathcal{O} \subset \mathbb{R}^n$ there is assigned a (not
necessarily unital) $*$-algebra $\mathcal{A(O)}$, and the condition of
isotony $\mathcal{O}_1 \subset \mathcal{O}_2 \Rightarrow
\mathcal{A(O}_1) \subset \mathcal{A(O}_2)$. Then one can form the
algebra $\mathcal{A}^{\circ} = \bigcup_{\mathcal{O}}\mathcal{A(O)}$
generated by all local algebras $\mathcal{A(O)}$, and we suppose that
$\mathcal{A(O)}^{\circ}$ is endowed with a locally convex topology in
such a way that it becomes a topological $*$-algebra.
We denote by $S_{\mathcal{A}^{\circ}}$ the set of all continuous
semi-norms on $\mathcal{A}^{\circ}$. 
 Moreover, we
suppose that there operates on $\mathcal{A}^{\circ}$ an
equi-continuous action of the translation group $(\alpha_x)_{x \in
  \mathbb{R}^n}$ fulfilling the condition of covariance,
$\alpha_x(\mathcal{A(O)}) = \mathcal{A(O} + x)$. (The condition of
equi-continuity says that for each $\sigma \in
S_{\mathcal{A}^{\circ}}$ there is $\sigma' \in
S_{\mathcal{A}^{\circ}}$ and $r > 0$ with $\sigma(\alpha_{x}(A)) \le
\sigma'(A)$ for $|x| < r$, and $\sigma(\alpha_x(A) - A) \to 0$ as $x
\to 0$ for each $a \in \mathcal{A}^{\circ}$.) There is yet another
condition concerning the structure of the $\mathcal{A(O)}$ that we
wish to impose here. Namely, we suppose that for each bounded open
region $\mathcal{O}$, it holds that $\mathcal{A(O)} = \bigcup_{k \in
  \mathbb{N}}\mathcal{V}_k(\mathcal{O})$ is the union of an ascending
sequence of vectorspaces $\mathcal{V}_k(\mathcal{O}) \subset
\mathcal{V}_{k+1}(\mathcal{O})$ with
$\alpha_x(\mathcal{V}_k(\mathcal{O})) \subset
\mathcal{V}_k(\mathcal{O} + x)$ for each $\mathcal{O}$ and $k$, with
$\mathcal{V}_k(\mathcal{O}) \cdot \mathcal{V}_{k'}(\mathcal{O})
\subset \mathcal{V}_{k''}(\mathcal{O})$ for some $k''$ depending on
$k$ and $k'$, and with the property that given $\sigma \in
S_{\mathcal{A}^{\circ}}$, $\mathcal{O} \subset \mathbb{R}^n$ open and
bounded, and $N,k \in \mathbb{N}$, there is some $\sigma' \in
S_{\mathcal{A}^{\circ}}$ so that 
$$
 \sigma(A_1 \cdots A_N) \le \sigma'(A_1) \cdots \sigma'(A_N)
$$ 
holds for all $A_1,\ldots,A_N \in \mathcal{V}_k(\mathcal{O})$.
\par
The just listed structural properties are typical of the Borchers
algebra, and will be made use of in order to incorporate the Borchers
algebra formulation of quantum field theory in our approach to
generalizing the wavefront set to the operator algebraic setting in
quantum field theory. However, the partitioning of local algebras into
subspaces $\mathcal{V}_k(\mathcal{O})$ would in general not be
regarded as an intrinsic element in the algebraic framework of quantum
field theory and is presently mainly to be seen as a technical device
in order to treat the Borchers algebra at equal footing with
$C^*$-algebras in our approach.
\par
Supposing the validity of the just stated assumptions, we define, for
each $0 \le \mu \le 1$, $p \in \mathbb{R}^n$ and for each bounded
open region $\mathcal{O} 
\subset \mathbb{R}^n$,  ${\bf A}^{(\mu)}_p(\mathcal{O})$
as the set of all families $(A_{\lambda})_{\lambda > 0}$ with the
following properties:
\begin{itemize}
\item[(i)] There is $k \in \mathbb{N}$ so that $A_{\lambda} \in
  \mathcal{V}_k(\lambda^{\mu}\mathcal{O} + p)$, $\lambda > 0$,
\item[(ii)] There is some $\lambda_0 > 0$ with $A_{\lambda} = 0$ for
  $\lambda > \lambda_0$,
\item[(iii)] For each $\sigma \in S_{\mathcal{A}^{\circ}}$ there is $s
  \in \mathbb{R}$ so that 
$$ \sup_{\lambda}\,\lambda^s \sigma(A_{\lambda}) < \infty \,.$$
\end{itemize}
The elements in ${\bf A}^{(\mu)}_p(\mathcal{O})$ are called
\textit{testing families}. Note that ${\bf A}^{(\mu)}_p(\mathcal{O})$
inherits in a natural way a linear structure by defining algebraic
operations on testing families pointwise for each $\lambda$. (If each
$\mathcal{V}_k(\mathcal{O})$ is an algebra, then so is
${\bf A}^{(\mu)}_p(\mathcal{O})$.) Another natural operation is to
shift a testing family so that their localization properties change:
We may define $\alpha_x(A_{\lambda})_{\lambda > 0} =
(\alpha_x(A_{\lambda}))_{\lambda > 0}$, then
$\alpha_x({\bf A}^{(\mu)}_p(\mathcal{O})) = {\bf A}^{(\mu)}_{p +
  x}(\mathcal{O})$.  Equipped with that
notation, we can now introduce the following definition.
\begin{definition}
Let $\varphi$ be a continuous linear functional on
$\mathcal{A}^{\circ}$, and let $N \in \mathbb{N}$. For $0 \le \mu \le 1$
and $N \in \mathbb{N}$, we call an element
$$ (p_1,\ldots,p_N;\xi_1,\ldots,\xi_N) \in (\mathbb{R}^n)^N \times
((\mathbb{R}^n)^N \backslash \{0\}) $$
a \textit{regular directed point} of order $N$ and degree $\mu$ for
$\varphi$ provided that the following holds: There exists an open,
bounded neighbourhood $\mathcal{O}$ of the origin in $\mathbb{R}^n$,
an open neighbourhood $V_{(N)}$ of $(\xi_1,\ldots,\xi_N)$ in
$((\mathbb{R}^n)^N \backslash \{0\})$, and some $h \in
\mathcal{D}((\mathbb{R}^n)^N)$ with $h(0) \ne 0$ so that for each
$(A^{(j)}_{\lambda})_{\lambda > 0} \in
{\bf A}^{(\mu)}_{p_j}(\mathcal{O})$
one has
$$ \sup_{\underline{k} \in V_{(N)}}\, \left| \int
  \textrm{e}^{-i\lambda^{-1} \underline{k}\cdot \underline{y}}
  h(\underline{y}) \, \varphi(\alpha_{y_1}(A_{\lambda}^{(1)}) \cdots
  \alpha_{y_N}(A_{\lambda}^{(N)}))\, d^ny_1 \cdots d^ny_N \right| =
O(\lambda^{\infty}) $$
as $\lambda \to 0$. Here, $\underline{k} = (k_1,\ldots,k_N)$ and
$\underline{y}= (y_1,\ldots,y_N)$ denote $N$-tupels of vectors in
$\mathbb{R}^n$ and correspondingly, $\underline{k}\cdot \underline{y}$
denotes the sum of the scalar products $k_j \cdot y_j$, $j =
1,\ldots,N$.
\par
The set of all regular directed points of order $N$ and degree $\mu$ of
$\varphi$ is denoted by $\textrm{reg}^{(N,\mu)}(\varphi)$. The
complement of that set in $(\mathbb{R}^n)^N \times 
((\mathbb{R}^n)^N \backslash \{0\})$ is
 denoted by $\textrm{ACS}^{(N,\mu)}(\varphi)$ and
will be called the \textit{asymptotic correlation spectrum} of
$\varphi$ of order $N$ and degree $\mu$.
\end{definition}  
Before adding a few remarks about the definition of the asymptotic
correlation spectrum, we give an example which ought to illustrate why
this notion may be viewed a generalization of the wavefront set. Take
as local algebras the sets $\mathcal{A(O)} = \mathcal{D(O)}$, with the
pointwise multiplication of functions, and
$\mathcal{V}_k(\mathcal{O}) = \mathcal{D(O)})$ for all $k \in
\mathbb{N}$. Let $u \in \mathcal{D}'(\mathbb{R}^n)$. In this case one obtains:
\begin{lemma}
$$ {\rm ACS}^{(1,\mu)}(u) = {\rm WF}(u)  \quad \textrm{for\ all}\
0 \le \mu \le 1\,.$$
\end{lemma}
The proof of this statement is easily obtained by a simple variation
of Prop.\ 2.1 in \cite{Ver.acs}.
\par
In \cite{Ver.acs}, we have introduced the algebras of testing families
only for the case $\mu =1$. This was strongly inspired by the ``scaling
algebra'' approach to the analysis of short distance behaviour in
quantum field theory introduced in \cite{BV1,BV2}. The dependence on
$\mu$ that has been added here is but one possible way of
generalization. It is obvious that ${\bf A}^{(\mu)}_p(\mathcal{O})$
becomes larger as $\mu$ decreases, and so one gets
$\textrm{ACS}^{(N,\mu)}(\varphi) \subset \textrm{ACS}^{(N,\mu')}(\varphi)$
for $\mu > \mu'$. The case $\mu = 0$ is in some way distinguished from the
cases $\mu > 0$. Let us consider this the case $\mu = 0$ under the
assumption that the $\mathcal{A(O)}$ are $C^*$-algebras, and
$\mathcal{V}_k(\mathcal{O}) = \mathcal{A(O)}$ for all $k$. Then the
condition that $(p_1,\ldots,p_N;\xi_1,\ldots,\xi_N) \in
\textrm{reg}^{(N,0)}(\varphi)$ can be formulated in the following way:
There is a conic open neighbourhood $\Gamma$ of $(\xi_1,\ldots,\xi_N)
\in ((\mathbb{R}^n)^N \backslash \{0\})$, an open neighbourhood
$\mathcal{O}$ of the origin in $\mathbb{R}^n$ and some $h \in
\mathcal{D}((\mathbb{R}^n)^N)$ with $h(0) \ne 0$, so that there is for
each $R \in \mathbb{R}_+$ some $C_R > 0$ with
$$ \sup_{A^{(j)}}\,\sup_{\underline{k} \in \Gamma}\, (1+
|\underline{k}|)^R \left|\int \textrm{e}^{-i \underline{k} \cdot
    \underline{y}}h(\underline{y})\,\varphi(\alpha_{y_1}(A^{(1)})\cdots \alpha_{y_N}(A^{(N)})) \,d^ny_1 \cdots d^ny_N \right| < C_R $$
where supremum is formed over all $A^{(j)} \in \mathcal{A}(\mathcal{O}
+ x_j)$ with $||\,A^{(j)}\,|| \le 1$, $j = 1,\ldots,N$. (Such a
definition of regular directed points has been suggested to the author
by K.\ Fredenhagen.) In other words, elements in
$\textrm{reg}^{(N,0)}(\varphi)$ are simultaneously and uniformly
regular directed points (in the sense of not being contained in the
wavefront set) of all distributions $\varphi_{A^{(1)},\ldots,A^{(N)}}$
given by $\varphi_{A^{(1)},\ldots,A^{(N)}}(y_1,\ldots,y_N) =
\varphi(\alpha_{y_1}(A^{(1)})\cdots \alpha_{y_N}(A^{(N)}))$; therefore
one obviously has 
$$ \textrm{ACS}^{(N,0)}(\varphi) \supset \textrm{closure}\,\left[
\bigcup_{A^{(j)}} \textrm{WF}(\varphi_{A^{(1)},\ldots,A^{(N)}})\,\right]\,,$$
but there is no assertion if the reverse inclusion holds.
\par
As may be expected, the basic properties of
$\textrm{ACS}^{(N,\mu)}(\varphi)$ are similar to those of the wavefront
set. For instance, $\textrm{ACS}^{(N,\mu)}(\varphi)$ is a closed subset
of $(\mathbb{R}^n)^N \times ((\mathbb{R}^n)^N \backslash \{0\})$ and
conic in the $\xi_j$. For a proof and some more discussion, see Prop.\
3.2 in \cite{Ver.acs}.
\par
We shall provide a few more examples. 
Let $\mathcal{B}$ denote the Borchers algebra over $n$-dimensional
Minkowski spacetime. For $\mathcal{O}$ a bounded open region in
$\mathbb{R}^n$, we define the subspaces $\mathcal{V}_k(\mathcal{O}) =
\mathbb{C} \oplus_{m = 1}^k (\otimes^m \mathcal{D(O)})$ of
$\mathcal{B}$ and the local algebras $\mathcal{A(O)} = \bigcup_{k=
  1}^{\infty}\mathcal{V}_k(\mathcal{O})$. The action of the
translations $(\tau_xf)(y) = f(y -x)$ on test-functions lifts to an
equi-continuous group action $(\alpha_x)_{x \in \mathbb{R}^n}$ by
automorphisms to $\mathcal{A}^{\circ}$ (note that 
$\mathcal{A}^{\circ} = \mathcal{B}$). Define the sets of testing
families ${\bf A}^{(\mu)}_p(\mathcal{O})$ with respect to these data,
and let $\omega_m$, $m \in \mathbb{N}_0$ denote the $m$-point
functions of a state $\omega$ on $\mathcal{A}^{\circ}$. Then it is not
difficult to check that 
\begin{equation}
\label{llll}
{\rm WF}(\omega_m) \subset {\rm ACS}^{(m,\mu)}(\omega)\,.
\end{equation}
In general, one won't expect equality to hold here, since with the
just given definition the ${\bf A}^{(\mu)}_p(\mathcal{O})$
are algebras which are quite large. However, if we restrict the choice
of $\mathcal{V}_k(\mathcal{O})$ to $\mathcal{V}_k(\mathcal{O}) =
\mathcal{D(O)}$ for \textit{all} $k$, and if ${\bf
  A}^{(\mu)}_p(\mathcal{O})$ is defined accordingly, then equality
holds in \eqref{llll}.
\par
Another example arises from quantum field theories on Minkowski
spacetime in the $C^*$-algebraic framework described in Sec.\
2. Suppose that we are given such an algebraic quantum field theory,
$(\{\mathcal{A(O)}\}_{\mathcal{O}\subset\mathbb{R}^4},(\alpha_x)_{x\in\mathbb{R}^4},\omega^{\circ})$.
Define ${\bf A}^{(\mu)}_p(\mathcal{O})$ with respect to
$\mathcal{V}_k(\mathcal{O}) = \mathcal{A(O)}$ for all
$k$. Furthermore, let $\omega$ be a state on $\mathcal{A}$ which is
induced by a $C^{\infty}$-vector for the energy, $\psi$, in the vacuum
GNS-Hilbertspace $\mathcal{H}^{\circ}$ (so that $\omega(A) = \langle
\psi,\pi^{\circ}(A)\psi\rangle$). We recall that $\psi$ is
$C^{\infty}$ for the energy if $\psi \in {\rm dom}((P_0)^s)$ for all $s
> 0$ where $P_0$ is the energy-operator (generator of the
time-translations) in $\mathcal{H}^{\circ}$. In this situation, we
obtain
\begin{theorem}
It holds that
$$ {\rm ACS}^{(N,\mu)}(\omega) \subset \Gamma^{\circ}_N $$
where $\Gamma^{\circ}_N$ is the set of all
$(p_1,\ldots,p_n;\xi_1,\ldots,\xi_N) \in 
((\mathbb{R}^4)^N \backslash \{0\})$ 
with the folowing properties: There exists a
non-directed graph $G_N$ with $N$ vertices, and with all pairs of
distinct vertices connected by exactly one directed edge and its
inverse, together with an immersion $\iota(\,.\,)$ of $G_N$ into
Minkowski-spacetime, where each curve $\gamma_{ij}$ is a straight geodesic
line segment (which may degenerate to a point if $\iota(v_i) =
\iota(v_j)$), at least one of which is causal, so that
$$ p_i = \iota(v_i)\,, \quad \xi_i = \sum_j \iota(\vec{e}_{ij}).
 $$
\end{theorem}
The proof of this theorem may be inferred by combining Thm.\ 4.6 in
\cite{BFK} with Thm.\ 5.1 in \cite{Ver.acs}.
\par
This shows that in flat spacetime, the spectrum condition places an
upper bound on ${\rm ACS}^{(N,\mu)}(\omega)$ which is of the form of a
microlocal spectrum condition, and in turn shows that the asymptotic
correletion spectrum may serve as a generalization of the wavefront
set in the operator algebraic framework. There is a further result in
support of this point of view: Suppose that one has, in the vacuum
representation of the algebraic quantum field theory 
$(\{\mathcal{A(O)}\}_{\mathcal{O}\subset\mathbb{R}^4},(\alpha_x)_{x\in\mathbb{R}^4},\omega^{\circ})$,
a scalar quantum field (Wightman field) $\mathcal{D}(\mathbb{R}^4)
\owns f \mapsto \Phi(f)$ affiliated to the local von Neumann algebras
$\mathcal{R(O)}$ and denote by
$$ \omega_N(f_1 \otimes \cdots \otimes f_N) = \langle \psi,\Phi(f_1)
\cdots \Phi(f_N) \psi \rangle $$
the $N$-point distributions correponding to a state $\omega$ induced
by a unit vector $\psi$ in the domain of the quantum field. In this
case one obtains an analogue of \eqref{llll}, namely
$$ {\rm WF}(\omega_N) \subset {\rm ACS}^{(N,\mu)}(\omega)\,.$$
We refer to \cite{Ver.acs} for a proof and further discussion. 
\section{Quantum Field Theory on Curved Spacetime, Encore}
We have seen that the asymptotic correlation spectrum appears as a
viable generalization of the wavefront set in the operator algebraic
approach to quantum field theory in Minkowski spacetime. The next step
consists in generalizing the notion of asymptotic correlation spectrum
to quantum field theory in curved spacetime. To this end, we are again
faced with the difficulty that there is no counterpart of the
translation group acting by isometries on a curved spacetime, since the
translation group played a significant role in formulating conditions
on the regular directed points. Nevertheless, it appears that the
basic idea underlying the definition of regular directed points of a
functional may be suitably generalized so as to cover also the
situation where the spacetime manifold possesses no non-trivial
isometries.
We will consider that situation at the end of this section.
\par
First, we will focus at the situation where some isometries are still
present. While a more general investigation of the asymptotic
correlation spectrum for general group actions is on the way
\cite{SVW}, we will here restrict attention to the simplest case. We
assume that there is a smooth, one-parametric group
$\{\theta_t\}_{t\in\mathbb{R}}$ acting by isometries on the globally
hyperbolic spacetime $(M,g)$. Moreover, we assume that its generating
vector field $X = \frac{d}{dt}\theta^*_t$  
is time-like and future-pointing. Thus $(M,g)$ is
\textit{stationary}. By ${\sf N}_X = \{(p,\xi) \in T^*M :
\xi(X_p)=0\}$ we denote the \textit{co-normal bundle} of $X$.
\par
Then let $\{\mathcal{A(O)}\}_{\mathcal{O}\subset M}$ be a net of
topological $*$-algebras on $M$ satisfying the asumptions of the
previous section with obvious changes. Furthermore, we suppose that
there is an equi-continuous group action $(\alpha_t)_{t\in\mathbb{R}}$
by automorphisms on $\mathcal{A}^{\circ}$ which is covariant with
respect to $(\theta_t)_{t \in \mathbb{R}}$, i.e.\
$\alpha_t(\mathcal{A(O)}) =\mathcal{A}(\theta_t(\mathcal{O}))$. The
definition of the sets ${\bf A}^{(\mu)}_p(\mathcal{O})$ of testing
families $(A_{\lambda})_{\lambda > 0}$, for
$\mathcal{O}$ an open neighbourhood of $p \in M$, is similar as in the
last section except that the localization condition is replaced by
\begin{itemize}
\item[(i')] There is $k \in \mathbb{N}$ so that $A_{\lambda} \in
  \mathcal{V}_k({\rm exp}_p(\lambda^{\mu}{\rm exp}^{-1}_p(\mathcal{O})))$,
  $\lambda > 0$,
\end{itemize}
where ${\rm exp}_p$ is the exponential map at $p$ (and it is
understood that $\mathcal{O}$ is in the domain of ${\rm exp}_p^{-1}$).
\par
Now let $\varphi$ be a continuous linear functional on
$\mathcal{A}^{\circ}$. We define:
\begin{definition}
${\rm reg}^{(N,\mu)}_{(\alpha_t)}(\varphi)$ is defined as the set of
all $N$-tuples of covectors\\
 $(p_1,\xi_1;p_2,\xi_2;\ldots;p_N,\xi_N)
\in ((T^*M)^N \backslash ({\sf N}_X)^N)$ with the following property: There
is an open neighbourhood $W_{(N)}$ of $(p_1,\xi_1;\ldots;p_N,\xi_N)
\in ((T^*M)^N \backslash ({\sf N}_X)^N)$, and there are open neighbourhoods
$\mathcal{O}_j$ of $p_j$ $(j = 1,\ldots,N)$ and a function $h \in
\mathcal{D}(\mathbb{R}^N)$ with $h(0) \ne 0$, so that for all
$(A^{(j)}_{\lambda})_{\lambda > 0} \in {\bf
  A}^{(\mu)}_{p_j}(\mathcal{O}_j)$ it holds that
$$
\sup_{(\underline{p}',\underline{\xi}') \in W_{(N)}}\,
\left| \int {\rm e}^{-i \lambda^{-1} \underline{t}\cdot
    \underline{\xi}'(X)}h(\underline{t})
  \varphi(\alpha_{t_1}(A^{(1)}_{\lambda}) \cdots
  \alpha_{t_N}(A^{(N)}_{t_N}))\, dt_1 \cdots dt_N \right| =
O(\lambda^{\infty})$$
as $\lambda \to 0$. We have written $\underline{t} = (t_1,\ldots,t_N)$
and $(\underline{p}',\underline{\xi}') =
(p'_1,\xi'_1;\ldots,p'_N,\xi'_N)$, and in the phase-factor
$\underline{t} \cdot \underline{\xi}'(X) = \sum_{j = 1}^N t_j \cdot
\xi'_j(X_{p_j})$. The set ${\rm ACS}^{(N,\mu)}_{(\alpha_t)}(\varphi)$
is now defined as the complement of ${\rm
  reg}^{(N,\mu)}_{(\alpha_t)}(\varphi)$ in $((T^*M)^N \backslash ({\sf
  N}_X)^N)$. 
\end{definition}
With this definition, ${\rm ACS}^{(N,\mu)}_{(\alpha_t)}(\varphi)$ is a
closed conic subset of $((T^*M)^N \backslash ({\sf N}_X)^N)$. The following
result is taken from \cite{SV1}, to which the reader is referred for a
proof.
\begin{theorem}
Let $\omega$ be a continuous state on $\mathcal{A}^{\circ}$ and assume
that $\omega$ is a ground state, or a KMS-state at inverse temperature
$\beta >0$ or the action of $(\alpha_t)_{t\in\mathbb{R}}$. Then it
holds that ${\rm ACS}^{(2,\mu)}_{(\alpha_t)}(\varphi)$ is either
empty, or
$$ {\rm ACS}^{(2,\mu)}_{(\alpha_t)}(\varphi) = \{(p,\xi;p',\xi') \in
((T^*M)^2 \backslash ({\sf N}_X)^2): \xi'(X_{p'}) > 0\,, \ \xi(X_p) +
\xi'(X_{p'}) = 0\}\,.$$
\end{theorem}
As an application of this last result, let
$\{\mathcal{A(O)}\}_{\mathcal{O \subset M}}$ be the net of local
$C^*$-algebras constructed for the free scalar field fulfilling the
wave-equation \eqref{KGeqn} on $(M,g)$. This net carries an automorphic action
$(\alpha_t)_{t\in \mathbb{R}}$ which is covariant with respect to
$(\theta_t)_{t \in \mathbb{R}}$ \cite{Dim}. Suppose that $\omega$ is a
quasifree state on $\mathcal{A} = \mathcal{A}[K,\sigma]$ which is a
grund state or a KMS-state at inverse temperature $\beta > 0$ for
$(\alpha_t)_{t \in \mathbb{R}}$. Since the two-point function
$\omega_2$ of $\omega$ is a distributional solution of the
wave-equation in both entries, it holds by a general result that ${\rm
  WF}(\omega_2)$ must be contained in the set $\mathcal{N}$ of
null-covectors defined in \eqref{nullcov}.   
Combining this with the statement of the last theorem yields that 
$$ {\rm WF}(\omega_2) = \{(q,\xi;q',\xi')\ \in \mathcal{N}_- \times
\mathcal{N}_+ : (q,\xi) \sim (q',-\xi')\}\,,$$
i.e.\ that $\omega_2$ is of Hadamard form \cite{SV1}. This generalizes
similar results which have been obtained, by other methods, for a
special class  of static
spacetimes in the ground state case \cite{FNW} and for static
spacetimes with compact Cauchy-surfaces in the KMS case
\cite{Jun}; the spacetimes covered by these previous works do not,
however, include some interesting situations like black holes while our
result does. Moreover,  the argument can be extended from the
particular example of a free scalar field to the case of vector (-bundle)
fields over $(M,g)$ satisfying a wave-equation and suitably
generalized versions of the canonical commutation relations, and to
Dirac-fields fulfilling canonical anti-commutation relations, in any
spacetime-dimension $\ge 2$ \cite{SV1,SV2}. It should also be noted
that ground states and KMS states as well as mixtures of such states
are passive states for which the 2nd law of thermodynamics holds
(i.e.\ one cannot extract energy from such states by cyclic
processes), cf.\ \cite{PW}. Thus, in the case of linear quantum fields obeying
wave-equations on stationary spacetimes, one can see that the
microlocal spectrum condition (or equivalently, the Hadamard
condition) is implied by passivity. This is further support to the
idea that the microlocal spectrum condition selects states which, in a
suitable sense, are dynamically stable.
 \par
Finally we shall give a -- tentative -- outline how one may proceed
in order to obtain a notion of asymptotic correlation spectrum in case
that the underlying spacetime manifold $(M,g)$ admits no non-trivial
isometries. In that case, there is no obvious definition of
Fourier-integrals of the form
\begin{equation}
\label{jake}
\int {\rm e}^{-i \lambda^{-1}\xi \cdot
  y}h(y)\varphi(\alpha_y(A_{\lambda}))\,d^ny
\end{equation}
that we have used above in testing the regularity of directions $\xi$
for a given functional $\varphi$ upon letting $(A_{\lambda})_{\lambda
  > 0}$ range through a collection of testing families suitably
localized at the base point $p$ to which the direction $\xi$ is
affixed. But we may think of the expression \eqref{jake}
as
\begin{equation}
\label{jamie}
\varphi(A(\lambda^{-1}\xi,\lambda))\,,
\end{equation}
i.e.\ the functional $\varphi$ tested by ``symbols'' of the form
\begin{equation}
\label{club}
A(\xi,\lambda) = \int {\rm e}^{-i\xi\cdot y}h(y)
\alpha_y(A_{\lambda})d^ny\,.
\end{equation}
One may then be inclined to take the right hand side of \eqref{club}
as a specific example of abstractly defined ``testing symbols''
$A(\xi,\lambda)$ which are characterized by a suitable asymptotic high
energy/short distance behaviour as would ensue for the right hand side
of \eqref{club}, if a spacetime-translation group action were present.
In other words, one may generalize the approach of the last chapter by
introducing suitable classes of ``testing symbols'' $A(\xi,\lambda)$
and by defining regular directions of $\varphi$ via the asymptotic
$\lambda \to 0$ behaviour of the quantities \eqref{jamie} for all such
testing families.
\par 
Then the question arises which conditions on the testing symbols
one should impose, and how to implement the just sketched idea. In the
remainder of this work, we shall make some suggestions towards that
question; however, we should warn the reader that these suggestions so
far haven't been tested in examples, and should be taken \textit{cum
  grano salis}. Our starting point is a net
$\{\mathcal{A(O)}\}_{\mathcal{O} \subset M}$ of $C^*$-algebras indexed
by the bounded open regions of some (globally hyperbolic) spacetime
$(M,g)$ (dim\,$M = 4$). This net is assumed to comply with
the conditions of isotony
and locality and moreover, it will be assumed that each
$\mathcal{A(O)}$ is a \textit{von Neumann} algebra acting on some
Hilbertspace $\mathcal{H}$. That is to say, we assume that some
Hilbertspace representation (or, equivalently, a suitable set of
states) has been chosen, and the basic approach is to provide a
definition of test-objects that allow it to decide if that
representation fulfills, in a suitably generalized sense, a
(microlocal) spectrum condition, in which case the representation may
be regarded as physical. 
\par
Let a point $p \in M$ be given, and let $\mathcal{O}$ be an open,
geodesically convex neighbourhood of $p$. We shall consider functions
$$ T^*\mathcal{O} \times (0,1) \owns (x,\xi;\lambda) \mapsto
A(x,\xi;\lambda) \in \mathcal{A(O)}$$
with the following properties:
\begin{itemize}
\item[(I)] In any coordinate system for $T^*\mathcal{O}$, and for all
  multi-indices $\alpha,\beta \in \mathbb{N}_0^4$, the weak partial
  derivatives 
 $$ D_x^{\alpha}D_{\xi}^{\beta}A(x,\xi;\lambda) $$
exist, are jointly (weakly) continuous in $x,\xi,\lambda$, and are
contained in $\mathcal{A(O)}$.
\item[(II)] In suitable coordinates, 
$$ \sup_{x \in K} \,\sup_{k \in V}
\,||\,\left.D_x^{\alpha}D_{\xi}^{\beta}A(x,\xi;\lambda)\right|_{\xi
  = \lambda^{-1}k}\,|| \le C_{K,V,\alpha,\beta} (1 + \lambda^{-1})^{m
  + |\alpha| - |\beta|} $$
hold for each compact subset $K \subset \mathcal{O}$ and each bounded
subset $V \subset T^*K$ with suitable constants $m,
C_{K,V,\alpha,\beta}> 0$. (This property is essentially what
characterizes operator-valued symbols in microlocal analysis, see
e.g.\ \cite{BWSch}.)
\end{itemize}
We collect all functions $A(\,.\,,\,.\,;\,.\,)$ with the just
described properties in a set denoted by ${\rm
  Sym}(p,\mathcal{O})$. We call it the set of \textit{testing symbols}
around $p$ localized in $\mathcal{O}$.
\par
Let us give an example of such testing symbols in a concrete case:
Take $p \in M$, and choose a coordinate system $(y^{\nu})$ with
 $y(p) = 0$ around
$p$. Let $f$ be a smooth test-function supported in a sufficiently
small coordinate ball around $p$, and define, in coordinates,
$f_{x,\lambda}(y') = f(y'/(\lambda^s)- x)$, where $s \ge 1$. Here, we
have identified $x$ with its coordinate expression $y(x)$. Denote by
$w_x^{(\lambda)}$ the Weyl-operator $\pi(W([f_{x,\lambda}]))$ in the
GNS-representation $\pi$ corresponding to a quasifree state on the
CCR-algebra of the Klein-Gordon field on $(M,g)$. Then a testing symbol is
obtained by setting
$$ A(x,\xi;\lambda) = \int {\rm e}^{-i\xi_{\nu}y^{\nu}}h(y)w_{x +
  y}^{(\lambda)}\,d^4y $$
for $x$ in a sufficiently small neighbourhood of $p$ and $h \in
\mathcal{D}(\mathbb{R}^4)$ having support sufficiently close to 0;
 the coordinates
used for $\xi$ are those induced by the chosen coordinate system.
\par
One can now introduce a notion of generalized asymptotic correlation
spectrum of order $2$ (the case of arbitrary order $N$ can be treated
similarly, we consider only $N =2$ for the sake of simplicity).
 Let $\omega$ be a state on $\mathcal{A}$, the
quasilocal algebra generated by the local von Neumann algebras
$\mathcal{A(O)}$, and let $p,p' \in M$. We say that $(p,\xi;p',\xi')
\in (T^*M \times T^*M) \backslash\{0\} $ is a {\it
  generalized regular directed point} of order $2$
 for $\omega$ if there are open
neighbourhoods $\mathcal{O}$ of $p$ and $\mathcal{O}'$ of $p'$, and 
an open neighbourhood  $W_{(2)}$ of
$(p,p';\xi,\xi') \in
(T^*M \times T^*M)\backslash \{0\}$, so that
$$ \sup_{ (x,k;x',k') \in W_{(2)}}
\,|\omega(A(x,\lambda^{-1}k;\lambda)A'(x',\lambda^{-1}k',\lambda))| = 
O(\lambda^{\infty}) \quad {\rm for} \quad \lambda \to 0$$
holds for all  testing symbols $A(x,k;\lambda) \in {\rm
  Sym}(p,\mathcal{O})$ and $A'(x',k';\lambda) \in {\rm
  Sym}(p',\mathcal{O}')$.  
Then ${\rm gACS}^{(2)}(\omega)$, the {\it generalized asymptotic 
  correlation spectrum of order 2} of $\omega$, is defined as the
complement in $(T^*M \times T^*M )\backslash \{0\}$
 of the set of all generalized regular
directed points of order 2 for $\omega$.
\par
From the example for testing-symbols above it is fairly plausible that,
in the case where 
the local von Neumann algebras $\mathcal{A(O)}$ are generated by a
quantum field $\mathcal{D}(M) \owns f \mapsto \Phi(f)$, and where
$\omega_2(f_1 \otimes f_2) = \omega(\Phi(f_1)\Phi(f_2))$ denotes the
corresponding two-point functions, one
should find
$$ {\rm WF}(\omega_2) \subset {\rm gACS}^{(2)}(\omega) \,.$$
However, in some sense the set of testing-symbols ${\rm
  Sym}(p,\mathcal{O})$ is too big: It is not related in any
obvious way to a ``dynamics'' of the quantum field theory given by the
net of von Neumann algebras $\{\mathcal{A(O)}\}_{\mathcal{O} \subset
  M}$. But then, on a general spacetime manifold it is not clear how a
notion of a dynamics is to be formulated. The approach which we
suggest is, therefore, that candidates for physical states should
``select'' their own (asymptotic) dynamics from the sets of testing
symbols: Consider the case that for some state $\omega$ on
$\mathcal{A}$, there are subspaces ${\rm
  Sym}_{\omega}(p,\mathcal{O}) \subset {\rm Sym}(p,\mathcal{O})$ such
that the generalized asymptotic correlation spectra ${\rm
  gACS^{(2)}_{\omega}}(\omega')$, defined with respect to ${\rm
  Sym}_{\omega}(p,\mathcal{O})$ instead of ${\rm Sym}(p,\mathcal{O})$,
have the property that, e.g.,
\begin{eqnarray*}
{\rm gACS}^{(2)}_{\omega}(\omega') \subset \{(p,\xi;p',\xi') \in
(T^*M \times T^*M) \backslash \{0\}&:& g^{\mu \nu}\xi_{\mu}\xi_{\nu} \ge
0,\ g^{\mu\nu}\xi'_{\mu}\xi'_{\nu} \ge 0,\\
 & & \xi(X) < 0,\ \xi'(X) >
0\}
\end{eqnarray*}
holds, for any timelike vector-field $X$, for a dense set of normal
states $\omega'$ (including $\omega$ itself). Then one would be
inclined to call such a state 
dynamically stable once the symbol-spaces ${\rm
  Sym}_{\omega}(p,\mathcal{O})$ are sufficiently stable under
algebraic operations like multiplication of symbols, or under the
convolution 
  $$ A \star A'(x,\xi,\lambda) = \int A(x,\xi
  -\xi',\lambda)A'(x,\xi',\lambda)\,d^4\xi' 
$$ 
in suitable coordinates; moreover, ${\rm Sym}_{\omega}(p,\mathcal{O})$
would have to be sufficiently ``big'' (e.g.\ ${\rm
  Sym}_{\omega}(p,\mathcal{O})'' = \mathcal{A(O)}$).
Further desiderata that one would like to impose on elements of ${\rm
  Sym}_{\omega}(p,\mathcal{O})$ are suitable (asymptotic) forms of geometric
modular action. These matters remain to be  explored; we just wished to
point out that it appears well possible to extend the microlocal
approach to generalizing the spectrum condition to quantum field
theory in curved spacetime in the operator-algebraic setting. It seems
also possible to further extend these ideas to reach at notions of
``spectrum condition'' for generally covariant theories, but this is
still quite speculative.
\small

\end{document}